# Active liquid crystals powered by force-sensing DNA-motor clusters.

Alexandra M. Tayar[1], Michael F. Hagan[2], Zvonimir Dogic[1,2*]

[1]Department of Physics, University of California, Santa Barbara, CA, USA.
[2]Department of Physics, Brandeis University, Waltham, MA, USA.

*Corresponding author: zdogic@physics.ucsb.edu

**Abstract:** Cytoskeletal active nematics exhibit striking non-equilibrium dynamics that are powered by energy-consuming molecular motors. To gain insight into the structure and mechanics of these materials, we design programmable clusters in which kinesin motors are linked by a double-stranded DNA linker. The efficiency by which DNA-based clusters power active nematics depends on both the stepping dynamics of the kinesin motors and the chemical structure of the polymeric linker. Fluorescence anisotropy measurements reveal that the motor clusters, like filamentous microtubules, exhibit local nematic order. The properties of the DNA linker enable the design of force-sensing clusters. When the load across the linker exceeds a critical threshold the clusters fall apart, ceasing to generate active stresses and slowing the system dynamics. Fluorescence readout reveals the fraction of bound clusters that generate interfilament sliding. In turn, this yields the average load experienced by the kinesin motors as they step along the microtubules. DNA-motor clusters provide a foundation for understanding the molecular mechanism by which nanoscale molecular motors collectively generate mesoscopic active stresses, which in turn power macroscale non-equilibrium dynamics of active nematics.

**Main text**:

**Introduction:** Active matter is composed of animate energy-consuming elements that collectively drive the system away from equilibrium, thus endowing it with life-like properties, such as motility and spontaneous flows (1). To make progress towards long-term applications of active matter, it is essential to elucidate the design principles for engineering large scale behaviors by controlling the dynamics of the microscopic constituents. Developing theoretical frameworks have the potential to describe diverse active matter systems, ranging from simple phase-separating colloidal swimmers to complex self-organized mitotic spindles (2–6). Importantly, many active matter systems have anisotropic constituents; thus, their theoretical description needs to account for the local orientational order (7, 8). In such materials, the locally aligned microscopic constituents generate internal active stresses, which in turn drive large-scale chaotic dynamics that includes autonomous flows and creation and annihilation of motile topological defects (9–16). An impactful class of active anisotropic fluids is based on reconstituted cytoskeletal elements, wherein the active stresses are generated by clusters of molecular motors that step along multiple filaments, driving their relative sliding (17, 18). So far, the focus has been on quantifying the chaotic dynamics of cytoskeletal active matter in both the nematic and isotropic phases, and methods of controlling their autonomous flows through boundaries and confinement (19–26). However, being reconstituted from well-defined biochemical components, these systems provide a unique, yet so far largely unexplored, opportunity to elucidate the microscopic origins of the emergent chaotic dynamics, thus paving the way for developing predictive multiscale models (27–29).

The key prerequisite for advancing this line of inquiry is the measurement of active stresses, as these are the primary generators of the non-equilibrium dynamics. This objective requires characterizing the force loads experienced by the molecular motors as they move in dense active nematic environments. Notably, there is a lack of studies in this direction, despite the force load being an essential determinant of motor stepping dynamics (30, 31). The challenge of characterizing these forces in active nematics is compounded by the fact that the motor clusters are advected by the rapid autonomous flows. Thus, measuring motor loads requires Lagrangian force sensors that move with the material. In contrast, most single-molecule techniques such as optical tweezers or atomic force microscopes work in the Eulerian coordinates, where the measurement spring is held fixed in the laboratory reference frame (32).

To address these critical questions, we take advantage of recent advances in DNA-nanotechnology that developed probes for quantifying forces in diverse biological processes, such as cell-matrix interactions, force-induced protein binding, and protein folding. DNA-based probes can be readily integrated into dynamical systems to report locally generated forces with a fluorescent reporter (33–37). Motivated by these results, we integrated DNA-based force-sensing kinesin clusters into microtubule-based active nematics. These clusters have a programmable binding strength, rupturing above critical applied stress, and thereafter ceasing to generate microtubule sliding and associated active stresses. Monitoring the fraction of bound clusters within an active nematic uncovers the average load experienced by the motors, an essential yet previously inaccessible parameter. With a few notable exceptions (38), so far almost all cytoskeletal active matter has been driven by irreversibly linked motor clusters. Programable DNA-based force-sensing motor clusters provide insights into both the molecular structure, mechanics, and active stress generation of microtubule-based active nematics.

Results

**Microtubule-based active nematics driven by DNA-kinesin clusters**: Conventional active nematics are assembled from stabilized microtubules (MTs), a non-absorbing depletant poly(ethylene glycol) (PEG), and streptavidin clusters of kinesin motors (17, 39). When sedimented on a surfactant-stabilized oil-water interface, the MTs form aligned nematics that are driven away from equilibrium by the motor clusters that convert energy from ATP hydrolysis into interfilament sliding motion. In conventional microtubule-based active matter, active stresses are generated by clusters of tetrameric streptavidin that binds together multiple biotin-labelled processive kinesin-1 motors with an exceedingly strong noncovalent bond.

To understand the mechanism of force generation, we assembled clusters in which two motors are linked by a hybridized double-stranded (ds) DNA (Fig. 1a, S1). Single-strands of DNA were modified at their 5'-ends with benzoguanine (BG), which formed a covalent bond with a SNAP-tag fused to the kinesin motor neck (40). DNA linker was formed from two complementary DNA oligos where the hybridized region ranged from 3 to 200 bp. Well-studied properties of DNA allow for the rational design of clusters whose motors were linked by a polymer of known length and elastic compliance. Furthermore, controlling the structure of the hybridized region yielded clusters that rupture above a critical pre-programed force. To understand the influence of the motor-stepping dynamics, we assembled clusters with two different kinesin motors (Fig. 1a,b). First, we used processive double-headed dimeric kinesin-1 motors. Two-headed kinesin takes about 100 consecutive 8 nm steps, before unbinding from a microtubule. At saturating ATP, each step takes ~10 ms (41). Second, we also assembled clusters of non-processive single-headed kinesin-1 motors. In contrast to processive kinesin, a single-headed motor attaches to a MT, takes a single step, and detaches (42–44).

We visualized active nematics by imaging either fluorescent MTs or the fluorescently labelled DNA clusters (Fig. 1c, S2). DNA-based clusters bind to neighboring microtubules. As they move towards MT plus ends kinesin motors generate interfilament sliding motion, which in turn generates large-scale chaotic flows. These flows are measured by embedding and tracking micron-sized passive tracer particles. Importantly, the samples exhibited a constant velocity for the duration of the experiment, and a spatially homogenous motor distribution (Fig. 1f). While processive and non-processive kinesins have markedly different single-molecule dynamics, they generated nearly identical collective behavior. We first examined the structural features of active nematics, before using the properties of DNA clusters to gain insight into mechanical forces experienced by motor clusters.

**Nematic order of DNA motor clusters:** Active nematics are characterized by the local orientational order of their MTs (Fig. 1c). In comparison, little is known about the alignment of motor clusters that generate the active stress. These could have an orientational order that ranges from nearly isotopic to perfectly aligned, and their alignment will impact the efficiency by which they generate dipolar extensile stresses. To gain insight into cluster orientations, we used fluorescence anisotropy measurements to estimate the orientation of the DNA linker. The intercalating fluorophores predominantly absorb and emit polarized light along their transition dipole moment. For the YOYO-1 fluorescent marker, this dipole is perpendicular to the DNA's long axis (45). We excited the DNA-clusters within the nematic film using polarized light, and measured the fluorescent signal that passed through an analyzer that was colinear with the incoming polarized light (Fig. 2a). Fluorescence anisotropy images showed a significant correlation between the local fluorescent intensity and the local MT orientation (Fig. 2b). In comparison, active nematics imaged with unpolarized light yielded a spatially uniform signal.

To estimate the nematic order of the DNA-clusters, we define $\phi$ as the angle between the DNA's long axis and the colinear polarizer/analyzer (P/A) (Fig. 2a). The emitted fluorescence along the P/A axis, $I_{||}$, is related to the excitation intensity $I$, as follows: $I_{||} = I \cdot \cos^2(90^0 - \phi)$. The phase shift is due to the fluorophore's dipole being perpendicular to the DNA's long axis. When $\phi = 90^0$, the dipole is parallel to the P/A axis, yielding a maximal signal. MT-based active nematics yielded maximum fluorescence when MTs were perpendicular to the P/A axis ($\phi = 90^0$), demonstrating nematic order of the DNA clusters, which are on average aligned along the microtubule axis. However, the alignment is far from perfect, as we still measure significant intensity when $\phi = 0^0$. We plot the fluorescence signal as a function of the local MT orientation with respect to the P/A axis, which is equal to the DNA alignment, $\phi$ (Fig. 2c). We assume that DNA-clusters locally have a Gaussian orientational distribution: $p(\phi) \propto e^{\frac{-(\phi-90^0)^2}{2\theta^2}}$, were $\theta^2$ is the variance. The measured fluorescent signal is given by: $I_{||} = \sum p(\phi) \cdot \cos^2(\phi - 90)$, where $p(\phi)$ is the fraction of motors pointing along $\phi$, and $\cos^2(\phi - 90)$ is the projection on the P/A axis. Using the variance as an adjustable parameter, we fitted the model predictions to experiments (Fig. 2c). The extracted variance is statistically different for different cluster types (Fig. S3). Subsequently, we used the fitted variance to estimate the nematic order parameter of the rod-like linker: $S = \langle \frac{3}{2}\cos^2\phi - \frac{1}{2}\rangle$. Long-linker clusters of processive motors (200 bp) yielded $S = 0.339 \pm 0.016$, while shorter ones (16 bp) had $S = 0.444 \pm 0.052$. In comparison, clusters with non-processive motors had an even higher order parameter of $S = 0.484 \pm 0.068$.

Very little is known about the microscopic structure of active nematics and how motors are arranged within a microtubule bundles, which makes it difficult to rigorously interpret the measurements of the order parameters. We found lower order for the long DNA clusters. Such clusters allow for a wider range of crosslinking conformation which could cause a wider distribution compared to short DNA. In active nematics, MTs have an almost perfect local alignment. Fluorescence anisotropy results suggest that kinesin clusters have significantly lower nematic order. Some caution is required when interpreting the fluorescence anisotropy experiments. In principle, there could be two distinct populations of motor clusters: one including the force-generating clusters with both motors attached to MTs, and the other including clusters with only a single bound motor. It is possible that the nematic order of the non-force-generating clusters is more isotropic than the doubly-bound clusters. Such bimodal distributions are not accurately described by the assumed Gaussian distribution. In this case, the measured signal, $I_{||}$, would overestimate the fraction of clusters that have a wider orientation. Thus, our analysis provides a lower bound estimate of the nematic order parameter for doubly linked clusters.

**DNA binding interactions control the dynamics in the active nematic film**: Next, we focus on studying the dynamics of active nematics and its dependence on both the DNA linker length and binding strength. In particular, DNA linkers allow for reversible assembly of two complementary strands, wherein the size of the hybridized region controls the binding strength. Such constructs can elucidate the minimum binding energy required to generate interfilament sliding. In this vein, we assembled clusters were hybridized regions ranged from 3 base pairs (bps) to 200 bps. For lengths up to 7 bps, thermal fluctuations alone break apart a measurable fraction of clusters in experimentally relevant temperature range. Beyond this limit, within the experimental error, essentially all clusters are permanently bound (Fig. S5). A hybridization length of 200 base pairs corresponds to a linker with a ~70 nm contour length. For physiological conditions the persistence length of DNA is 50 nm, so the longest linkers studied are semi-flexible filaments (46).

Depending on the binding energy, we identified three regimes of active stress generation (Fig. 3a). Clusters with a short-hybridized region (<3 bp) are not stable even in the absence of external load (Fig. S5). Consequently, in this regime, most clusters attach to MTs in the monomeric (unbound) state. Thus, they are unable to generate inter-filament sliding and active stresses, and there is no discernible motor-driven dynamics. In this *weak binding* regime, the MT networks are not fluidized and do not sediment to the oil-water interface to form an active nematic (Fig. 3b).

Increasing the hybridization length increases the binding energy and the fraction of bound motors. In this *optimal binding* regime, the hybridized region ranges from 7 to 32 bps and clusters primarily bind to MTs in paired form. Such clusters generate inter-filament sliding and active stresses, which leads to a robust dynamic that is faster than the background activity due to nonspecific motor aggregation (Fig. 3b, S4). Importantly, the velocity of the nematic flows increases with increasing DNA hybridization length (Fig. 3c, S6). In this regime, both processive and non-processive clusters exhibit the same qualitative behavior, while showing different velocities for different hybridization lengths.

Finally, in the *stretching regime*, hybridization lengths greater than 32 bps lead to irreversibly bound clusters, even in the presence of motor generated forces. Thirty base pairs corresponds to ~10 nm; hence the linker lengths are comparable to the kinesin step size (41, 43, 44). We hypothesize that the mechanism of active stress generation in this regime occurs in multiple steps. First, the cluster binds to two MTs, typically with its DNA linker having some slack and not oriented perfectly parallel to the MTs. Second, the motors need to take one or more steps, to fully stretch and orient the DNA linker. Clusters generate MT sliding and active stresses only once its linker is fully stretched (Fig. 3a). This hypothesis is supported by the marked differences observed for clusters of processive and non-processive motors. With increasing linker length, the dynamics of nematics powered by processive clusters reached a maximum velocity for $v_{dimer} \sim 2.5\ \mu m/s$, before decreasing slightly for longest linkers studied. Processive kinesin motors move continuously over ~1 μm distances. Hence, they are able to stretch clusters with long-linkers and generate active stresses. In comparison, for non-processive clusters the active nematics speed increased with linker length, and reached a peak velocity for 16 bp linkers. Beyond 21 bp, the velocity sharply decreased. In this regime, the non-processive kinesins are unable to stretch the cluster with a single step that is a few nanometers in size, hence there is a significant reduction in interfilament sliding and active stress generation (43, 44). Intriguingly, the spatial structure of the active nematics was largely not dependent of the nature of DNA clusters (Fig. S7).

**Quantifying the fraction of paired force-generating motors**: The above-described findings demonstrate that active stress generation requires paired clusters. Furthermore, the temporal stability of the autonomous dynamics suggests that the fraction of stress generating clusters remains constant (Fig. 1f, S8). To make progress, it is essential to quantify the fraction of paired clusters. Clusters that generate active stress are paired through DNA hybridization. Thus, quantifying the amount of ds-DNA within an active nematic will yield the fraction of motors capable of generating stress. We accomplished this by using SYBR-green, a dye whose fluorescence is both linearly dependent on the ds-DNA concentration and increases by a thousand-fold upon binding to ds-DNA. Using SYBR-green we label active nematics that are powered by DNA-clusters of single-headed motors (Fig. S9). In comparison to double-headed clusters, the simplified structure and lower background activity of the single-headed motor allow for quantitative measurements (Fig. 3c, S4) (47).

We first measured the fraction of paired clusters in equilibrium samples without molecular motors and microtubules using melting curves (Fig. 4a, S5). For 7 bp linkers, ~85% of the clusters were bound. Beyond this overlap length, at room temperature, essentially all DNA was hybridized, within the measurement error. Next, using confocal imaging we quantified the SYBR-green fluorescence in an active nematic powered by single-headed kinesin motors (Fig. 4a). The fraction of paired clusters was determined by normalizing the measured signal with the signal of active nematics powered by 16 bp bound clusters that are irreversibly bound, while accounting for the differences in the hybridization lengths. In active nematics, DNA-linkers are under tension generated by molecular motors and sliding MTs. Therefore, we hypothesized that the fraction of paired clusters would be reduced in active samples when compared to quiescent solutions described above. Indeed, we found that activity significantly reduced the fraction of bound clusters. For example, for 7 bp clusters activity decreased the fraction of paired clusters from 0.85 to 0.362 ± 0.036 while for 9 bp clusters the equivalent decrease is from 1.0 to 0.786 ± 0.063 (Fig. 4a).

Next, we verified the activity-induced decrease in the fraction of paired clusters using an independent measurement. Specifically, we compared the speed of the active nematics powered by reversible clusters that are continuously interconverting between paired and unpaired states, to samples containing a predetermined and known fraction of paired and unpaired clusters that cannot interconvert between each other. To accomplish this goal, we mixed motor clusters that are never paired (0 bp) with those that are irreversibly paired (16 bp). We then measured the active nematic speed as we changed the fraction of the two cluster types, while keeping the overall concentration constant. Measurements with these standardized samples yielded the same quantitative dependence of the velocity on the fraction of paired clusters as those obtained with the SYBR green method, thus validating the proposed method for estimating the fraction of bound clusters (Fig. 4b, c Red data set).

**Quantifying cluster binding to MTs within active nematics**: Quasi-2D active nematics assemble by depletion-induced adsorption of MTs from a 3D suspension onto a surfactant stabilized oil-water interface(48). While MTs are strongly adsorbed to the interface, all other components, including the motor clusters, can continuously exchange with the aqueous reservoir above the interface (Fig. 5a). Quantifying this exchange dynamics is essential for developing models of microtubule-based active nematics. To determine the partitioning of motor clusters between the 3D reservoir and 2D active nematic, we measured the $z$-dependent fluorescence signal using confocal microscopy. We used 16-bp clusters, in which a fluorophore is covalently attached to the DNA linker (Fig. 5b, S10). The measured signal was maximal in the nematic film, quickly decaying to a background constant value, which was ~25% of the maximum. We translate the fluorescence intensity into physical meaningful concentration units as follows. The concentration of the motors in the reservoir is 350 nM. The nematic layer thickness is estimated to be 120 nm (SI), which yields a concentration of motors in the layer of ~800 clusters/μm$^3$. From this, we estimate that there are 3 kinesin clusters (6 kinesin motors) per each microtubule (SI).

To quantify the exchange kinetics between the active nematics and the reservoir, we photo-bleached a ~20 μm$^2$ square area in a nematic assembled at lower ATP concentrations (100 $\mu M$) and measured the fluorescence recovery (Fig. 5c). For fluorescent MTs, the bleached area remained constant, indicating an absence of significant filament exchange on the time scale of minutes (Fig. 5d). Subsequently, we bleached fluorophore-labeled motor clusters. In principle, cluster fluorescence could recover by either exchange with the 3D reservoir or by motors stepping along the MTs from the unbleached borders. For both processive and non-processive clusters the signal recovered uniformly, demonstrating that the exchange with the 3D reservoir dominates the cluster dynamics. Notably, the fluorescence of both cluster types did not recover to their original values, suggesting that a fraction of motors remained MT-bound on longer time scales (Fig. 5d, S11). The fluorescence of processive motor clusters recovered to lower values compared to non-processive motors. Possible reasons include the lower exchange kinetics of such clusters are due to the effects of kinesin aggregation and formation of higher-order structures, the presence of rigor motors, and damage from photobleaching.

Fluorescence recovery curves yield the estimates of the effective rates at which motor clusters unbind from the nematic layer, $k_{\text{off}}$. When diffusion is faster than the molecular binding rate, the recovery of fluorescence is determined by $k_{\text{off}}$ (49). We fit the data to an exponential curve $\propto (1 - e^{k_{\text{off}} \cdot t})$ (Fig. 5e). The measured unbinding rate exhibited a weak dependence on cluster concentrations, for both processive ($k_{\text{off}} \sim 0.04 \ s^{-1}$) and non-processive clusters ($k_{\text{off}} \sim 0.1 \ s^{-1}$). A lifetime of 10 s for non-processive motors in the nematic film is significantly longer than the duration of a single step, which is a few milliseconds. This suggests that the depletion forces and highly crowded environment within 2D active nematics induce multiple consecutive MT-kinesin stepping events before a cluster dissociates from the nematic.

**Estimating the external load on DNA-motor clusters:** We developed a simple model to relate the activity-induced cluster unbinding to the average load experienced by the motor clusters. A paired motor cluster is under tension due to direct forces that are exerted as its motors step along MTs. There are also indirect forces on the cluster, due to other motors that slide the MTs apart as well as the associated hydrodynamic flows. In steady-state, the fraction of bound DNA that forms clusters is determined by the balance between the DNA unbinding rate ($k_{\text{off}}$) with the rate of DNA binding ($k_{\text{on}}$). We assume that activity primarily increases $k_{\text{off}}$ as motor generated forces shear the dsDNA that holds the cluster together. Additionally, we assume that activity-induced forces vanish for ss-DNA motor clusters; thus, $k_{\text{on}}$ is activity-independent. With these assumptions, the fraction of unpaired clusters can be predicted by estimating the load-dependent increase of $k_{off}$, without the explicit knowledge of the absolute value, $k_{off}$. Optical tweezer measurements quantified the dependence of dsDNA rupture force on the hybridization length (50). Combining these experiments with molecular simulations provides a quantitative model of how $k_{off}$ changes with the applied force (51). In particular, at steady-state, the force-dependent "binding constant", $K$, is given by the balance between on- and off-rates of the DNA:

$$K(N,f) = \frac{k_{on}}{k_{off}(N,f)} = K(N,0)e^{-\frac{(N\delta-\delta_0)f}{k_B T}},$$

where $N$ is the number of base-pairs in the hybridization region, $f$ is the applied force, $\delta$ is the extension (per base pair) of the DNA at the transition state, and $\delta_0$ is an offset that allows for some base pairs to remain intact at the transition state.

Our model predicts how the fraction of paired clusters depends on the hybridization length and the force across the DNA linker (Eq. S1-S4). With increasing force, the point where 50% of clusters are paired shifts to larger hybridization lengths (Fig. 6a). This can be quantitatively compared to the experimental measurements (Fig. 4a). As the hybridization length is experimentally controlled, the only free parameter is the applied force. The force load that yields optimal agreement with experiments is $f^* = 2.9 f_{stall}$, where $f_{stall} \approx 7$ pN is a load for which the kinesin velocity decays to zero (31). Thus, our model implies that motors operate close to or even above their stall loads. We emphasize that the above-described estimate makes simplifying assumptions. Specifically, details such as the specific sequence of the hybridized region and the ionic strength will shift the exact location of the optimal binding regime.

**Relating the velocity profile to DNA-linker properties:** Next, we model the dependence of the active nematics speed on the structure of the DNA linker and the motor processivity (Eq. S5). We first estimate the number of DNA clusters for which the attached motors are actively pulling pairs of neighboring microtubules (Sup Info). We assume that the force generation requires: (1) paired clusters, (2) two motors that are attached to a pair of antiparallel microtubules, and (3) alignment of clusters with the MTs so that motor-generated forces predominantly induce microtubule sliding rather than cluster reorientation. To calculate the MT sliding speeds, we use a mean field estimate of the relationship between the density of the active motors, which assumes a linear force-velocity relationship for motors (Sup info). We assume that processive motors, on average, move $l_{\text{proc}} =$ 800 nm before unbinding, while non-processive motors take a single 8 nm step. Motivated by the possibility that an effective processivity arises due to depletant-induced attractions between the motor constructs and microtubules, we have also considered an intermediate processivity length of 10 steps.

The calculated interfilament sliding velocity as a function of DNA linker length and motor processivity (Fig. 6b) exhibits similar trends as the experimental observations (Fig. 3c). Note that the magnitude of the measured speeds differs by orders of magnitude. Our model predicts the relative sliding speed of two neighboring filaments. The filaments are extending everywhere within the active nematics. The mechanisms by which these local extension generate much faster large-scale dynamics are described elsewhere (52). The predicted decrease in velocity with hybridization for low processivity clusters is more gradual in our model when compared to experimental observations. A possible reason is the assumption that the rate of motor construct reorientation is limited only by the kinesin stepping rate, whereas motor construct motions could be impeded by the dense environment of the active nematic.

**Discussion:** Active nematics are powered by kinesin clusters that simultaneously bind to two antiparallel MTs. However, the microscopic details of how motors power interfilament sliding are unknown. We demonstrated that single-headed non-processive motors power active nematics as efficiently as processive motors. Furthermore, our analysis reveals that the force load on single-headed motor clusters is ~20 pN, which is significantly larger than the 7 pN stall force that has been measured for processive motors. At first, these findings might appear inconsistent with the efficient generation of interfilament sliding, which requires that both motors are simultaneously engaged with two MT filaments. Analysis of conventional motility assays powered by single-headed kinesins suggests that motors are engaged with a MT at most ~50% of the time (44). Thus, the probability of both motors being simultaneously engaged with MTs seems small. Naively one would expect that the external load would enhance the motor unbinding and thus further reduce the efficiency of interfilament sliding. However, recent studies demonstrated that the kinesin-MT unbind rate is highly dependent on both the direction and the magnitude of the external load, suggesting a possible mechanism that resolves above-described inconsistencies (53, 54). In particular, resisting loads applied along the MTs long axis significantly *decrease* the kinesin-MT unbinding rate. For example, a ~20 pN resisting load increases the kinesin-filament bond lifetime by multiple orders of magnitude when compared to load-free conditions. In comparison, forces perpendicular to the MTs long axis decrease the bond lifetime. Thus, for certain conditions, kinesin forms a catch bond whose strength increases with the applied load.

In active nematics, the two motors are coupled via a linker, which ensures that they experience resisting loads. Furthermore, the nematic alignment of clusters demonstrates that the resistive loads primarily point along the MTs long axis, a direction that maximally increases the lifetime of the MT-kinesin bond. Thus, load-dependent unbinding might be essential for the efficient generation of interfilament sliding. Large loads increase the bond lifetime, which greatly increases the efficiency of clusters crosslinking two filaments and inducing their relative sliding.

Two types of events could drive cluster rupture. The clusters could rupture due to forces applied by the motors during their power stroke. Alternatively, the clusters could rupture while passively linking a filament pair whose relative sliding motion is powered by other motors. The direct force produced by a motor construct occurs primarily during the power stroke of each kinesin, which has a ~10 μs time scale. This is a small fraction of the entire hydrolysis cycle, which at saturating ATP lasts ~10 ms (31). The probability of bond rupture depends on both the magnitude of the applied force and the time scale over which this force is applied (55). Because of its short duration, forces acting during the power stroke alone would have to be orders of magnitude larger to induce unbinding of a significant fraction of clusters. Instead, our results are consistent with cluster unbinding induced by forces that are on the order of, or larger than, the stall force, and are applied over a large fraction of the motor hydrolysis lifetime. Thus, forces experienced by the motor constructs arise primarily due to microtubule motions induced by other motors within the nematic.

In the conventional view, MT-based active nematics are viscous fluids in which motor clusters step along two microtubules, generating interfilament sliding that drives large-scale chaotic motion. Intriguingly, recent experiments visualizing single filament dynamics suggest that interfilament sliding in a dense nematic is not easily connected to the sliding of individual microtubule (52). The results described here suggest an alternative scenario. On average, there are 6 kinesin motors interacting with each filament, with each motor applying a ~20 pN force for the majority of its lifetime. Each cluster likely links different MT pairs. Thus MT-based liquid crystals are heavily crosslinked structures similar to previously studied gels linked with kinesin-14 motors (28). Individual motors attached to any given MT push in opposite directions. Thus, forces on a MT are mostly balanced, which gives rise to large pre-stress, as has been measured in actomyosin gels (56). In such materials, MT motion would arise from fluctuations in the net force. Multiple microscopic events could cause unbalance in the net force: (1) a motor might take a power stroke that increases the DNA linker tension, (2) a motor could unbind and release the tension, or (3) the DNA-linker could rupture. These results demonstrate the need to develop novel rheological techniques capable of characterizing 2D active nematics. Furthermore, having an estimate of load on each linker reveals that the average stresses exerted by the motors in the gel are ~1-40 kPa (SI).

In summary, we developed a programmable kinesin motor cluster capable of driving MT-based active nematics. The unique capabilities of the developed system provide new insight into possible mechanisms by which nanometer-sized kinesin motors drive macroscale chaotic flows. More broadly, our system illustrates potential synergies that arise by merging the precision of the DNA nanotechnology with the emerging field of active matter.

## Methods

**Tubulin purification and microtubules polymerization:** Tubulin was purified from bovine brain through 2 cycles of polymerization and depolymerization (57). Tubulin was stored at -80° C and subsequently recycled through an additional polymerization and depolymerization step. For fluorescent imaging, tubulin was labeled with Alex-647 dye (Invitrogen, A-20006) using a succinimidyl ester linker (58, 59). Absorbance spectrum showed that the percentage of labeled tubulin was 30-60%. Microtubules (MTs) were polymerized from a mixture of recycled tubulin and 4% labeled monomers in a buffer containing 10 mM GMPCPP (Jena Biosciences), 20 mM DTT in M2B (80 mM PIPES, pH 7, 1 mM EGTA, 2 mM $MgCl_2$). The final tubulin concentration was 8 mg/ml. The suspension was incubated at 35°C for 30 min, allowed to sit at room temperature for 5 hours, flash frozen in liquid nitrogen, and stored at -80° C.

**Kinesin purification:** Kinesin-401(dimeric kinesin) and kinesin-365 (monomeric kinesin) consist of 401 and 365 amino acids of the N-terminal motor domain of *D. melanogaster* kinesin. Both motors were cloned with fusion to the SNAP tag, and purified as previously published (60). The SNAP-tag is appended to the cargo binding region of the motor. The protein was flash frozen in liquid nitrogen and stored at -80° C.

**DNA-BG labeling:** 5'-amine modified DNA oligos (IDT) were labeled with BG-GLA-NHS (NEB) (40, 61). Briefly, oligos at 2 mM concentration were mixed with BG-GLA-NHS (15-20 mM in DMSO) in HEPES buffer (200 mM, PH 8.4) at a volume ratio of 1:2:3. BG-GLA-NHS was added last to the mixture. The mixture was incubated for 30 min at room temp. DNA was separated from excess BG using size exclusion spin column (Micro Bio-Spin 6 columns, Bio-Rad). Prior to DNA cleaning, tris-buffer in the column was exchanged with PBS (PH 7.2) according to manufacturer instructions. The separation step was repeated 4 times. The labeling efficiency, between 70-100%, was determined by DNA gel electrophoresis (20% TBE acrylamide gel for 60 min at 200 V). Labelled DNA oligos were stored at -20° C.

**Assembly of DNA-motor clusters:** DNA oligos were annealed to their complementary strands. DNA mixture in a duplex buffer (100 mM Potassium Acetate; 30 mM HEPES, pH 7.5) was heated to 95° C for 10 min, and gradually cooled down to room temperature in a heat block left on the bench. Annealed DNA were either stored at -20° C or used immediately. ds-DNA with BG-modified 5' ends was mixed with SNAP-tagged kinesin motors at a molar ratio of 2:1 (kinesin:DNA). The DNA-kinesin mixture was incubated for 30 min at room temperature prior to the experiment. Dimeric kinesin has two heads for each DNA, whereas monomeric kinesin has one head. The formation of motor clusters was verified using gel electrophoresis, revealing that ~70-80% of the DNA was labeled with the kinesin, for both monomeric and dimeric kinesin (Fig. S1).

**MT-based active nematic:** Active nematics were assembled as described previously (62, 63). The only difference involved the assembly of flow chamber with rain-X treated coverslip instead of Aquapel.

**Visualizing DNA active nematics:** In experiments requiring visualization of DNA, MTs were not labeled to prevent signal blead-through. DNA was labeled with YOYO-1 (Fisher Scientific) at a concentration of 100-200 nM, or SYBR green at a dilution of 1:5000. For measuring *z*-profiles, DNA clusters were labeled internally with either Cy3 or Cy5 fluorophore.

Active nematics were imaged using conventional fluorescence microscopy using Nikon Ti-2 and an Andor-Zyla camera running open-source microscopy managing software Micro-Manager 1.4.23. DNA intercalators were imaged using scanning laser Leica-SP8 confocal microscope. To suppress light inhomogeneity throughout the sample and minimize signal from different z-sections, the polarization anisotropy imaging was conducted with a Leica-SP8 confocal microscope. A polarizer in the light path that served both as a polarizer and analyzer. Photobleaching recovery experiments were conducted on the SP8-Leica confocal with a 20x NA 0.75 objective and a 488 nm laser source at 40X zoom. Velocity measurements were conducted by particle tracking (Alexa-488 labeled silica beads 3 $\mu m$ diameter) or with the MatLab PIV tool.

**\*** This manuscript was primarily supported by the Department of Energy, Basic Energy Sciences through award DE-SC0019733. Development of the theoretical model was supported by NSF DMR-1855914 and the Brandeis Center for Bioinspired Soft Materials, an NSF-MRSEC (DMR-2011846). We also acknowledge use of the Brandeis MRSEC BioSynthesis Facility which is supported by DMR-2011846. Alexandra M. Tayar is a Simons Foundation Fellow of the Life Sciences Research Foundation and is an Awardee of the Weizmann Institute of Science -National Postdoctoral Award Program for Advancing Women in Science. We also acknowledge the KITP Active20 program, which is supported in part by the National Science Foundation under Grant No. NSF PHY-1748958. Illustrations were created with BioRender.com

**Figures**

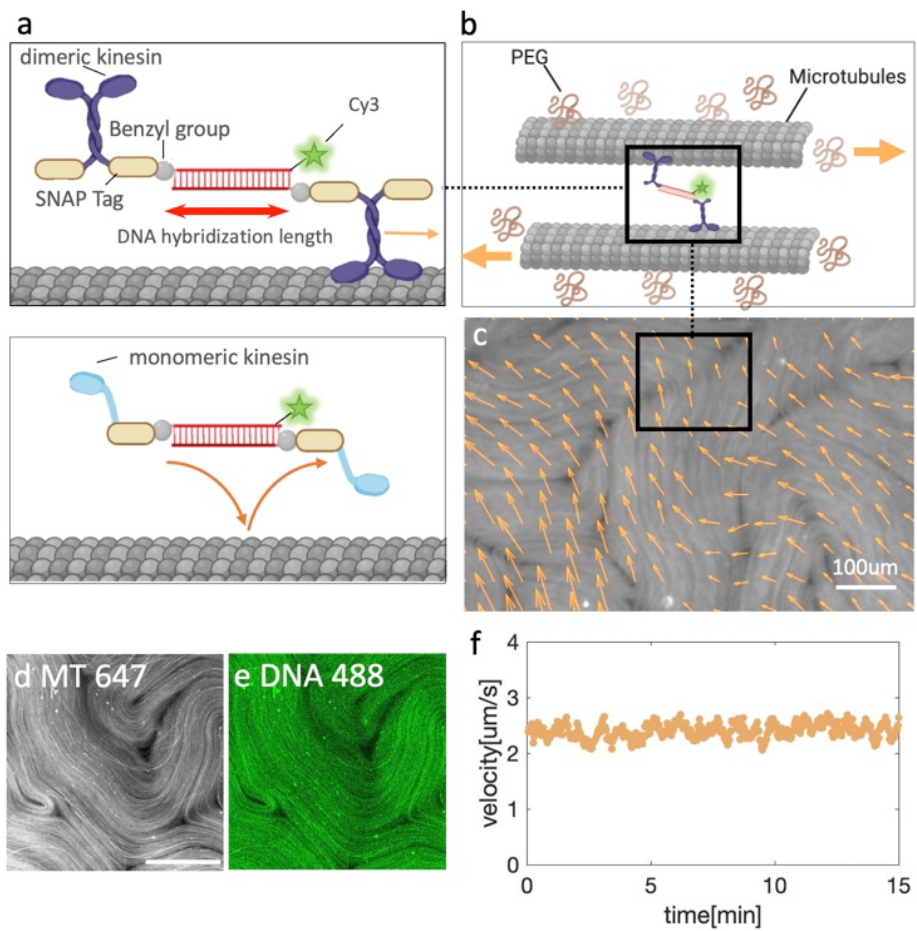

**Fig. 1- Active nematics powered by DNA-kinesin clusters. (a)** Schematic of a DNA motor cluster. Processive (double-headed) and non-processive (single-headed) kinesin motors bind to the dsDNA's end by a SNAP-BG covalent bond. The DNA is internally labeled with a fluorophore. The hybridization length controls the cluster binding strength. Double-headed kinesins moves processively on MTs for ~100 consecutive steps. Single-headed kinesins unbind from MTs after each step. **(b)** Schematic of a kinesin-MT bundle, the elemental structural motif that exerts extensile stresses and drives the active nematic. MTs are bundled by the depletion agent PEG, and motor clusters crosslink the filaments and induce their sliding. Bundles are confined to a surfactant-stabilized oil-water interface, where they form a dense 2D nematic film. **(c)** Fluorescence image of a 2D active nematic film, microtubules labeled; arrows indicate local velocity magnitude and direction. **(d)** Active nematic composed of fluorescent microtubules. **(e)** Active nematic containing unlabeled MTs but fluorescently labelled DNA clusters. Scale bar, 100 $\mu m$. **(f)** Spatially average velocity of autonomous flows of an active nematic film showing stability over time. The velocity is measured from micron-sized passive tracer particles embedded in the layer averaged over space.

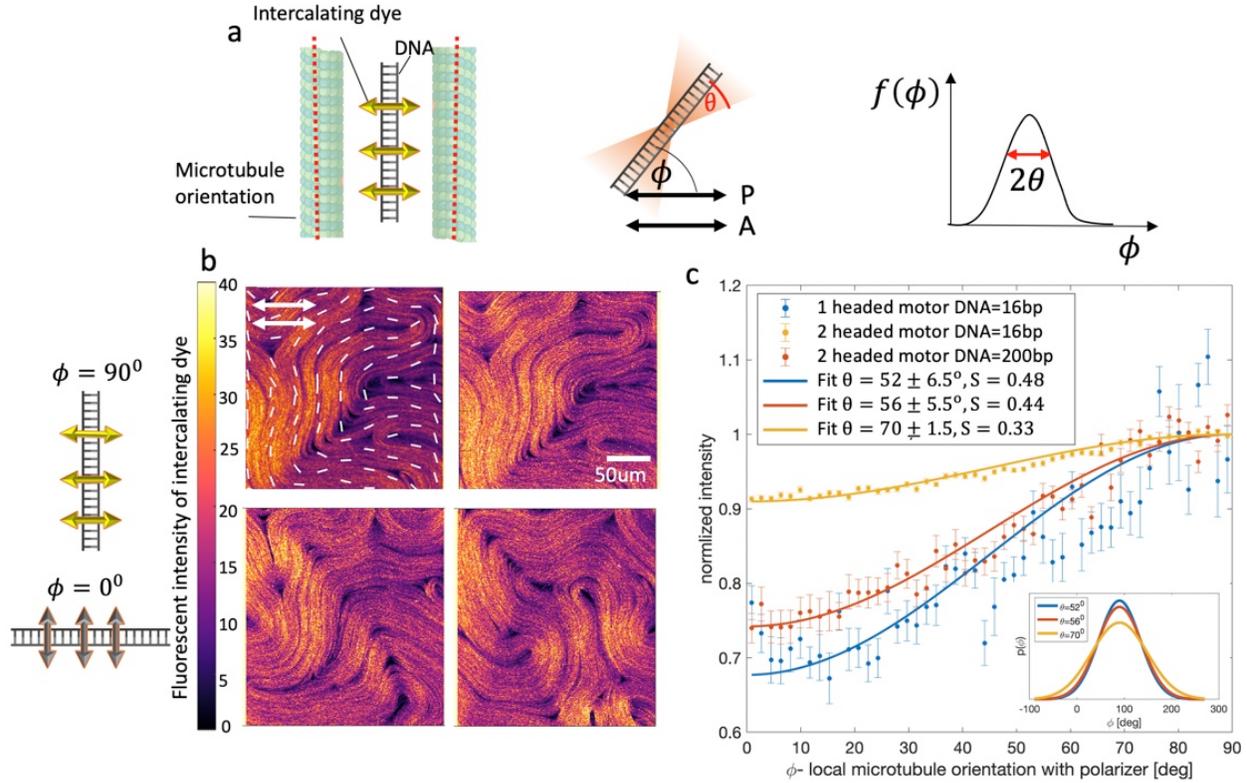

**Fig. 2 - Orientational order of DNA clusters (a)** Schematic of a DNA motor cluster labeled with an intercalator dye orientation between two MTs. Polarized light excites the intercalated dyes, and the emitted fluorescence is measured using a colinear analyzer. Orientation of intercalating dye dipoles indicated with bidirectional arrows; the dye molecules are at $90^0$ to the DNA long axis. $\phi$ is the angle between the DNA long axis and the polarizer/analyzer (P/A) axis. $\theta$ is the standard deviation of the DNA orientational distribution. **(b)** Fluorescence anisotropy of active nematics. Bidirectional arrows indicate P/A orientations. White lines correspond to director field. Emitted fluorescence is maximum when dye points along the P/A axis. **(c)** Normalized fluorescence intensity as a function of $\phi$ for 3 different cluster types: 100 bp double-headed clusters, 15 bp double-headed clusters, and 15 bp single-headed clusters. Maximal intensity at $\phi = 90^0$ indicates that the DNA is aligned with the MTs. Error bars are standard errors averaged over $n = 20\text{-}100$ events in a measured angle. Lines are fit to a model assuming a normal distribution of angles $\phi$. The values of $\theta$ are estimated by least square method minimizing $\chi^2$, the errors are estimated by increasing $\chi^2$ in 1 from the minimum value, giving a variation of one standard deviation in $\theta$. Inset: probability distribution, $p(\phi)$ for the three different cluster types.

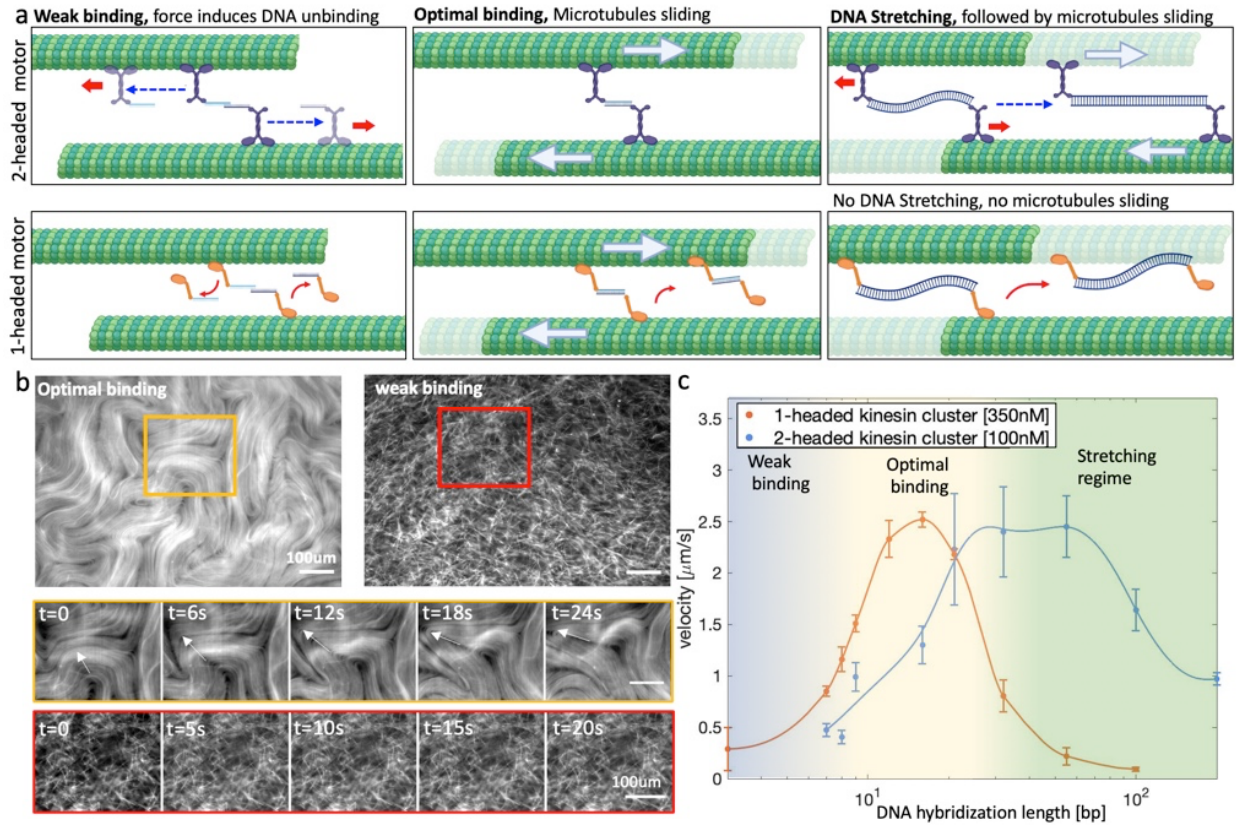

**Figure 3- Cluster binding energy controls active stress. (a)** Changing the DNA motor cluster hybridization length reveals different regimes of active stress generation. *Weak binding:* for short hybridization lengths (< 6 bp), the cluster binding energy is a few $k_BT$; hence clusters are unable to generate interfilament sliding. *Optimal binding:* for intermediate hybridization length 7-21bp, the strong cluster binding energy enable motor drive interfilament sliding. *DNA stretching:* for long hybridization lengths (> 21 bp) processive motors take multiple steps to stretch the DNA and then generate interfilament sliding. Single-headed clusters are unable to stretch the linker in a single step; thus, they generate no sliding. **(b)** An active nematic in the optimal binding regime (16 bp), and an isotropic static network formed in the weak binding regime for a single headed motor (3 bp). **(c)** Average speed of active nematic flows as a function of the linker hybridization length, for both processive (blue) and non-processive (orange) clusters. Velocities are measured from tracking 3 $\mu m$ beads embedded in the nematic. Error bars are standard errors over N=4-8 measurements.

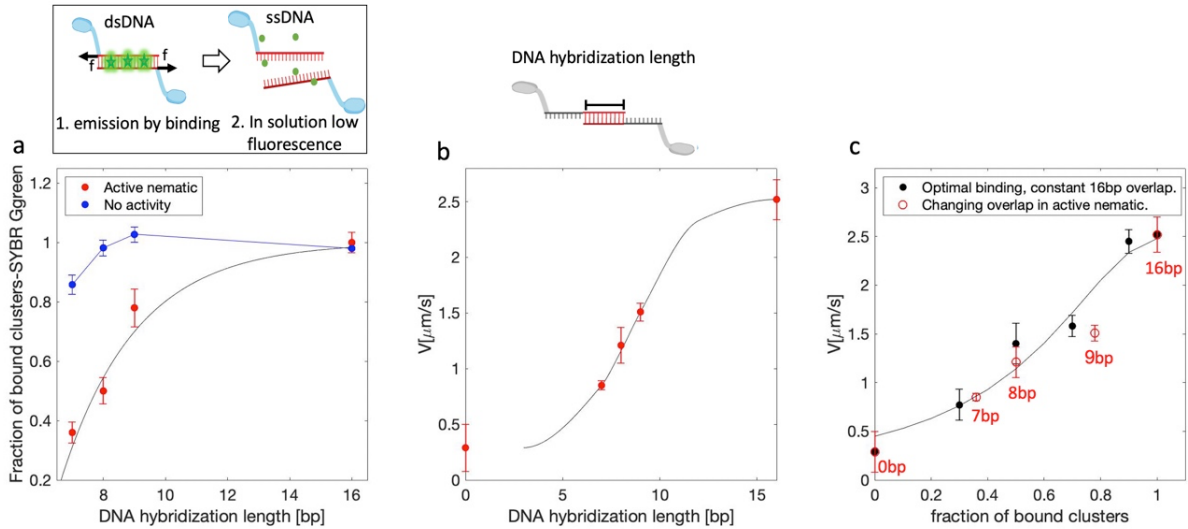

**Figure 4- Quantifying the fraction of paired motor clusters. (a)** Fraction of bound clusters as a function of DNA hybridization length in active nematics (red dots) and equilibrium suspensions (blue dots). Error bars are standard errors over 8-10 samples. **(b)** Speed of active nematic flows as a function of hybridization length. **(c)** Active nematic speed as a function of fraction of paired clusters (black points). The fraction of bound motors was controlled by changing DNA hybridization length; the data is taken from the y-axis value of the red data points in panels a and b. (red points) Control experiments where the fraction of paired and unpaired clusters was determined by changing the ratio of 16 and 0 hybridization length linkers. Error bars are standard errors over 6-8 samples.

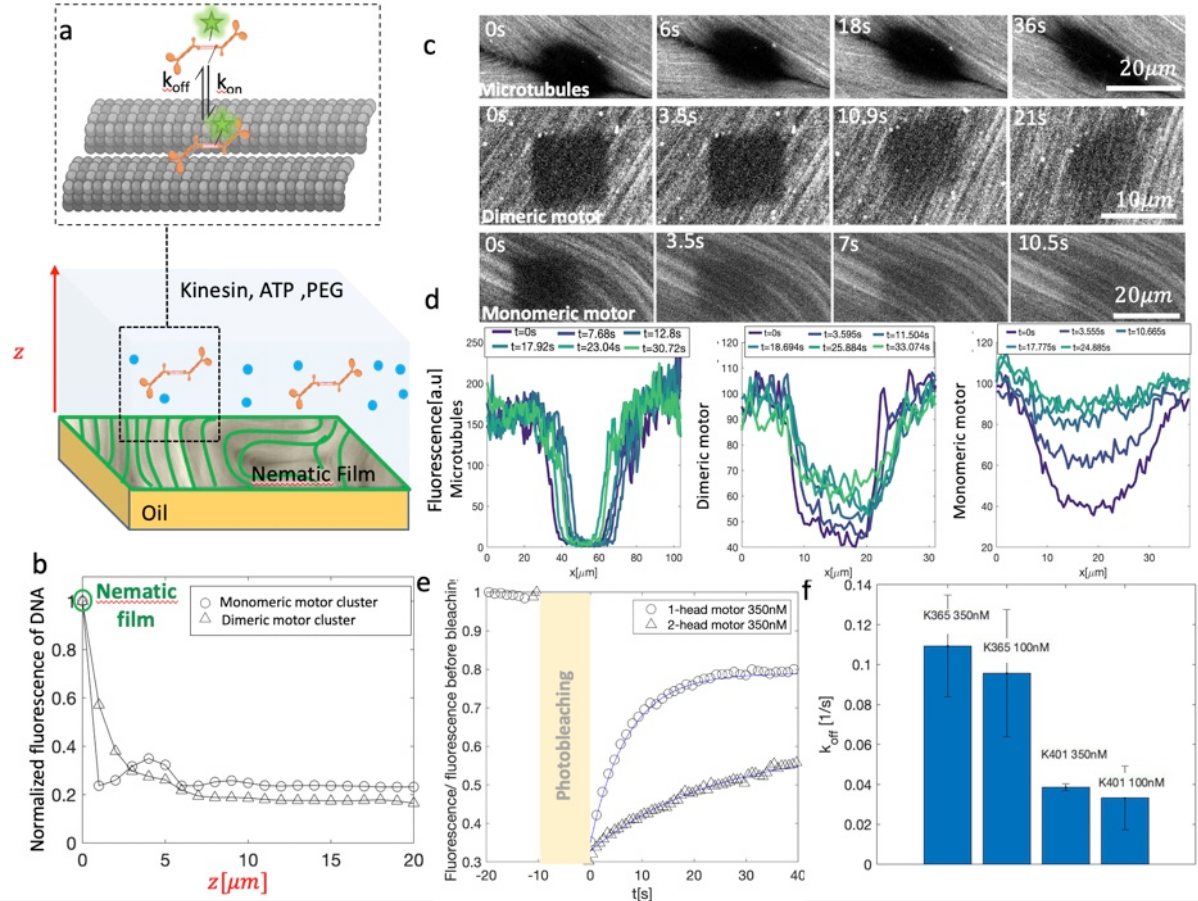

**Figure 5- Binding of motor clusters to 2D active nematics. (a)** 2D active nematic assembled on a surfactant-stabilized oil-water interface. Clusters partition between 2D nematics and the 3D reservoir located on the aqueous side of the interface. **(b)** Z-dependent fluorescence intensity profiles of the motor clusters. The nematic is located at z=0. **(c)** Fluorescence recovery after photobleaching (FRAP) images of fluorescently labeled MTs, processive, and non-processive motor clusters. DNA motor clusters are covalently labeled. **(d)** Spatial profile along the bleached window for MTs, processive, and non-processive motor clusters. **(e)** A Temporal profile of the fluorescence recovery after bleaching. Lines are fits to: $a \cdot (1 - be^{-k_{\text{off}} \cdot t})$. **(f)** $k_{\text{off}}$ for processive and non-processive motor clusters. Errors are the standard deviation (n=4). FRAP experiments were conducted at $[ATP] = 100$ μM to slow the system dynamics.

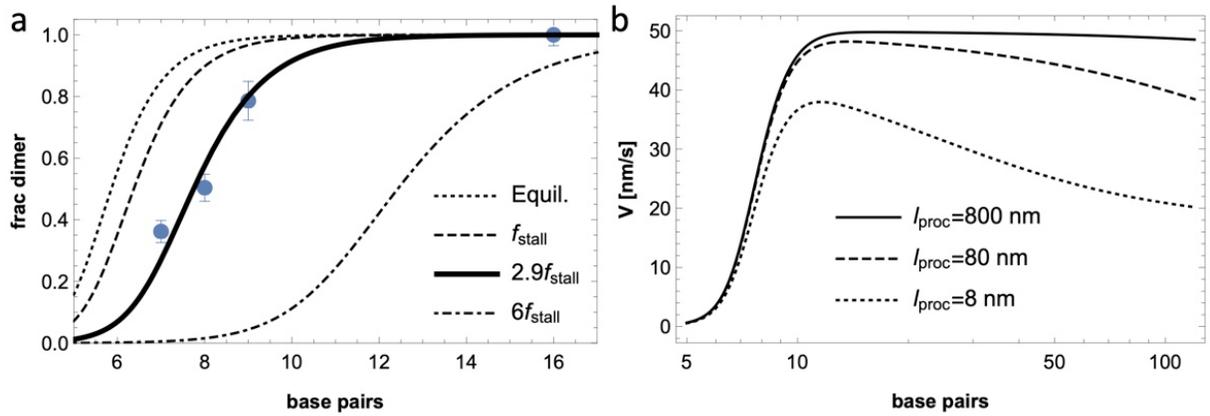

**Fig. 6- Force load determines the fraction of paired clusters. (a)** Fraction of hybridized dimers as a function of hybridization length in equilibrium and under different force loads. Experimental data is indicated by circles. The best fit to experimental data is obtained for $f = 2.91\, f_{stall}$, where $f_{stall} = 7$ pN. **(b)** Relative velocity between MT pairs predicted as a function of base pair overlap for processivity lengths corresponding to 1, 10, and 100 steps, with a step length of 8 nm. Error bars are standard error over 3-4 measurements.

**Supplementary information for Active liquid crystals powered by force-sensing DNA-motor clusters**

**Active nematics powered by processive vs. non-processive motors**: Dimeric kinesin forms higher-order aggregates due to nonspecific interactions. This leads to measurable non-equilibrium dynamics even in the absence of any specific motor linkages (Fig. S5). Monomeric kinesis offers a number of advantages. Only one BG label per motor interactions generates clusters with a better-defined number of motors. Moreover, monomeric motors have simplified dynamics as they detach from the MTs after each step, preventing passive crosslinking of the network (1) (Fig. 1a).

To test for the presence of non-specific oligomerization, we measured the activity of active nematics powered by processive and non-processive motors in the absence of any linking DNA. Background speeds were measured as a function of the kinesin concentration (Fig. S4). Nematics powered by processive motors exhibited a linear dependence of nematic speeds on the motor concentration, whereas non-processive motor clusters showed low background speeds of ~0.2 μm/s with a weak dependence on the motor concentration. All data was acquired for motor concentrations where the difference between velocity caused by non-specific interactions and velocity of DNA motor clusters was maximal. For monomeric kinesin the cluster concentration was 350 nM. For dimeric motors, the motor cluster concentration was 100 nM. Notably SNAP-tagged processive kinesins have two binding sites for BG labeled DNA and can form high-order multimeric DNA clusters. To minimize this possibility, we adjusted the ratio of DNA to kinesin such that for every two kinesin dimers there was one dsDNA linker. The DNA linkers were saturated in kinesin making assembly of high-order clusters less probable.

**Spatial distribution of motor clusters:** We measured spatial fluorescent profiles of the DNA and microtubules and plotted local microtubule intensity as a function of motor intensity. For both 16 bp and 8 bp overlaps, the local cluster density scaled linearly with the microtubule density and was independent of the local curvature (Fig. S2). Thus, we do not observe spatial variations in motor concentration that correlate with position along the defect, and we see no indication of heterogenous forces.

**Estimating motor generated active stresses:** To estimate the average number of kinesin clusters bound to each MT, we first estimate the nematic film thickness. A single MT measured in an aqueous suspension has a retardance area $R_0 = 7.5 \text{ nm}^2$, defined as the image retardance integrated along the axis perpendicular to the filament. It is independent of optical conditions such as focus, magnification, and numerical aperture (NA) (2). Additionally, the retardance of a MT bundle increases linearly with the filament number. The retardance of a nematic layer assembled under the conditions described here is $r \sim 2$ nm (3). Densely packed MTs with high concentration of depletant are ellipses spaced by 18.5 nm on their short axis and 34 nm apart on their long axis, giving a cross-section area of $A \sim 500 \text{ nm}^2$ (4). Therefore, from dimensional analysis, the thickness of the nematic layer can be estimated as $d = \frac{r}{R_0/A} \sim 120$ nm (2).

Next, we estimate the average number of kinesin clusters bound to a MT. For the single-headed kinesin, the bulk concentration of motors is 350 nM, corresponding to 200 DNA clusters/µm³. The concentration measured from a confocal optical volume of 1 µm³ focused on the nematic layer is roughly 4 times the concentration in the bulk, corresponding to ~800 clusters/µm³ (Fig 5). Further, we assume that all of the 800 clusters in the measured focal volume are bound to the 120 nm thick nematic layer, neglecting the freely diffusing clusters infinitesimally close to the layer. Therefore, the local concentration in the nematic layer is $6.6 \cdot 10^3$ clusters/$\mu m^3$. A nematic layer has a volume fraction of 0.12 X 1 X 1 $\mu m^3$, a microtubule has a cross section of 500 $nm^2$ and an average length of 1 $\mu m$, therefore a nematic layer contains ~240 MTs. This order of magnitude estimate indicates about ~3 clusters on each MT. Our work reveals that each of these clusters on average experiences a 20 pN force. From here, we can estimate the prestress generated by the motor clusters in a 3D film from the number of clusters cutting a plane perpendicular to the nematic layer of $0.12 \times 1$ µm².

As an upper bound on the stress, we assume that the MTs act as rigid rods on the micron scale, and thus are under a force on the order of the estimated force on each motor, 20 pN. Given that there are about 250 MTs cutting across a cross-section of 0.12 µm², the stress is $\sigma \sim 20 \cdot 250/0.12 = 4 \times 10^4 \frac{\text{pN}}{\mu m^2} = 40$ kPa. Alternatively, we obtain a lower bound on the stress as follows. We assume that forces from different motors are incoherent, and thus the motor force acts only over a short distance along the MT. Considering the effective size of a motor to be its step-length, 8 nm, on average this section will contain ~6.4 motor clusters. From here it follows that the stress is $\sigma \sim 1000 \frac{\text{pN}}{\mu\text{m}^2} = 1$ kPa.

**Model of fraction of hybridized dimers within the active nematic:** To estimate how the active stresses induce cluster unbinding, we developed a simple theoretical model. Our calculation is motivated by the observation that for clusters with base pair numbers ($N \leq 8$), the population of intact dimers in the active nematic is significantly lower than in the nematic without activity. We lack a detailed understanding of this environment to explicitly calculate the expected force, and we do not know hybridization kinetics within the dense environment of the nematic. Instead, we perform a calculation that does not require explicit knowledge of hybridization rates.

Within the active nematic, the DNA dimers will be subjected to direct forces from their attached motors, as well as other forces due to surrounding MTs and hydrodynamic flows. Thus, activity could contribute to dimer unbinding in multiple ways. To gain insight into this complex environment, we start with the assumption that the primary effect arises from motor forces shearing the double-stranded DNA molecule that holds the motor dimer together, thus increasing the cluster unbinding rate, $k_\text{off}$. We assume that this force vanishes for monomer (non-hybridized) motor constructs; thus, the hybridization on-rate of motor is force independent. The steady-state fraction of paired clusters is then equal to the value at which the force-enhanced off-rate matches the force-independent on-rate. Importantly, this population can be estimated based on the amount by which force *increases* off-rates, without explicit knowledge of the off-rates themselves.

We estimate the increase in the off-rates from experimental data on force-induced rupture of DNA dimers. This data was interpreted using a molecular computational model (5, 6). In this model, unbinding of a DNA molecule with $N$ base pairs is estimated from the extent by which a force $f$ applied along the length of the DNA reduces the transition barrier height for base pair unbinding:

$$\frac{k_\text{off}(N,f)}{k_\text{off}(N,f=0)} = \exp[(N\delta - \delta_0)f/k_\text{B}T], \qquad (1)$$

where $\delta$ is the extension (per base pair) of the DNA molecule at the transition state, $\delta_0$ is an offset that allows for some base pairs to remain intact at the transition state, and $k_\text{B}T$ is the thermal energy. Fitting unbinding rates using available experimental data estimated $\delta \approx 0.14$ nm and $\delta_0 \approx$ $\delta$, implying that approximately one base pair remains intact at the transition state (5, 6). The conditions for the active nematics in our work were not the same as those used in the single molecule experiments (5). Thus, we only expect qualitative agreement.

At steady-state, the force-modified "equilibrium constant" is given by the balance between on- and off- rates:

$$K(N,f) = \frac{k_{\text{on}}}{k_{\text{off}}(N,f)} = K(N,0)e^{-\frac{(N\delta-\delta_0)f}{k_BT}} = \frac{c_2(N,f)}{c_1^2(N,f)} \qquad (2)$$

where $K(N,0)$ is the true equilibrium constant (zero force) with $N$ base-pairs. Finally, the total concentration of motor clusters $c_T$ is related to the concentrations of monomers ($c_1$) and dimers ($c_2$) by mass conservation as $c_T = c_2 + \frac{c_1}{2}$, so the fraction of hybridized molecules, $P_2(f) = c_2(f)/c_T$, is given by:

$$P_2(N,f) = \frac{1}{8c_TK(f)}\left[1 + 8c_TK(N,f) - \sqrt{1 + 16c_TK(N,f)}\right] \qquad (3)$$

We estimate the equilibrium constants $K(N,0)$ from the hybridization free energies obtained from (7) (Table S1). To predict dimer fractions at arbitrary overlap, we fit the hybridization free energy $G(N)$ as a function of number of base pairs $N$, resulting in:

$$\frac{G(N)}{k_BT} = 1.7 - 2.56\,N. \qquad (4)$$

Eqs. (1-4) predict the fraction of bound dimers as a function of base pairs and force.

**Structure of DNA clusters determines active nematic velocity:** As a first step toward estimating the filament velocity, we calculate the fraction of the paired DNA clusters within the active nematic. We assume that active pulling requires paired clusters, whose two motors are attached to a pair of antiparallel MTs, and that the clusters are sufficiently aligned with the filaments so that forces primarily displace the MTs rather than reorient the clusters. Specifically, we consider a paired cluster whose motors bind to a pair of antiparallel MTs with a mean initial angle $\theta_0$ relative to the displacement vector between the two attached motors (Fig. S12). As the motors walk along their respective MTs, their applied forces simultaneously reorient the associated cluster while laterally displacing the MTs toward each other. MTs and paired clusters are roughly parallel after each of the two motors walks a distance $l = \frac{L_{DNA}}{2}(1 - \cos(\theta_0))$ along its MT, with $L_{DNA} = N \times 0.34$ nm the contour length of the cluster linker. Before that point, only a component of the motor force will displace MTs relative to each other. We assume the following form for the fraction $f_{frac}$ of the force magnitude $f_{motor}$ directed along the MTs as a function of angle $\theta$:

$$f_{frac}(\theta) = \frac{f(\theta)}{f_{motor}} = \begin{cases} 0 \text{ for } \theta \geq \frac{\pi}{2} \\ \cos\theta \text{ for } \frac{\pi}{2} > \theta \geq 0 \end{cases}$$

where $f(\theta)$ is the force along the microtubule at angle $\theta$. For simplicity, we assume that cluster unbinding kinetics is uncoupled from its reorientation. Furthermore, while the experiments indicate that on average motor constructs are more aligned with MTs than indicated by a uniform distribution, this measurement could be heavily skewed by constructs which are already bound to MTs. In particular, the population of motor clusters that is not bound to the DNA likely has a distribution closer to isotropic than measured, which would also correspond to the distribution of angles for clusters that have initially bound to MTs. Therefore, we assume the analytically tractable case of a uniform distribution of initial angles between 0 and $\pi$. The results will not be qualitatively affected by changing the initial distribution.

With these considerations, the fraction of force exerted by dimers within the active nematic that are actively displacing MTs is given by:

$$f_{active} = \frac{P_2(N,f)}{N} \int_0^\pi d\theta_0 P(\theta_0) \left[ \int_0^{\frac{L_{DNA}}{2}(1-\cos(\theta_0))} dl\, f_{frac}(\theta(l)) P_{motor}(l) \right.$$

$$\left. + \int_{\frac{L_{DNA}}{2}(1-\cos(\theta_0))}^\infty dl\, P_{motor}(l) \right]$$

with $\cos\theta = l/L_{DNA}$, and

$$P_{\text{motor}}(l) = \exp\left(-\frac{2l}{l_{\text{proc}}}\right)$$

the probability that both motors remain bound after walking a distance $l$, with $l_{\text{proc}}$ the processivity length of the motor, and the distribution of initial binding angles uniform as noted above,

$$P(\theta_0) = 1/\pi \quad \text{for } \theta_0 \in [0, \pi).$$

The normalization constant is given by:

$$N = \int_0^\pi d\theta_0 P(\theta_0) \int_0^\infty dl\, P_{\text{motor}}(l) = \frac{l_{\text{proc}}}{2}.$$

and we obtain

$$f_{\text{active}}(N, f) = \frac{1}{\pi} + \frac{1}{2L_{\text{eff}}}[1 + (1 - 2e^{-L_{\text{eff}}})I_0(L_{\text{eff}}) - \mathbf{L}_0(L_{\text{eff}})]$$

with $L_{\text{eff}} = L_{\text{DNA}} / l_{\text{proc}}$, $I_\alpha$ the modified Bessel function of the first kind, and $\mathbf{L}_\alpha$ the modified Struve function.

Finally, we calculate the filament velocity based on a mean field estimate of the relationship between the density of active motors (8–11), assuming a linear force-velocity relationship for motors(12–14) (although other relationships are possible depending on cross-linking density and degree of alignment (15))

$$v_{\text{filament}}(N) = \frac{2v_{\max}\rho_{\text{eff}}f_{\text{active}}(N,f^*)f^*}{v_{\max}\Gamma_\parallel + \rho_{\text{eff}}f_{\text{active}}(N,f^*)f^*}. \quad (5)$$

with $\rho_{\text{eff}} \sim c\rho_m$ the linear density of motors on MTs multiplied by a cooperativity factor accounting for the fact that motors will interfere above some density, $\Gamma_\parallel$ the longitudinal friction constant (per unit length) for a MT, and the factor of 2 in the numerator accounting for the fact that the two filaments are each moving away from each other. We set the maximum motor velocity to that of kinesin, $v_{\max} = 600$ nm/s, and the MT friction constant (per MT length) to $\Gamma_\parallel \approx$ $\frac{2\pi\eta}{\log(L/d_{MT})} = 1.3 \times 10^{-3}$ Pa·s, with $\eta = 1$ cp the viscosity of water and the MT aspect ratio $L/d_{MT} = 80$ with $L = 2\mu m$ the MT length. We assume that the dimers are on average subjected to the force $f^*$ determined above, but that they can only pull MTs with the maximum force $f_{stall}$. This approximation seems justified based on the fact that the maximum measured filament velocity $v_{filament}^{max} \approx 50$ nm/s is far below the maximum motor velocity. Within this limit, $v_{filament} \ll v_{max}$, the filament velocity is approximately given by $v_{filament}(N) \approx 2\rho_{eff} f_{active}(N, f^*) f^* / \Gamma_\parallel$. We also assume that the dimerization fraction $P_2(N)$ is independent of processivity length, because it was not experimentally possible to measure dimerization fractions for the processive K401 motor constructs. However, in reality we expect the average force on the processive motor constructs to be larger, and hence the dimerization fractions to be smaller at low overlap lengths. We discuss the effects of this assumption below.

The calculated filament velocity is a function of $N$ and motor processivity length (Fig. 6b). Here we have fit the unknown effective density $\rho_{eff} = 4.8 \times 10^{-6}$ $\mu m^{-1}$ by requiring that the calculated velocity match the measured filament velocity $v_{filament} \approx 50$ nm/s for the highly processive K401 in the high base pair limit, where $P_{active} \approx 1$. Given that the concentration of clusters in the film corresponds to a linear density of ~3 clusters/$\mu m$ (if all clusters are bound to microtubules), the extremely small value of this quantity indicates a relatively high effective friction due to motor cross-links and/or a lack of cooperativity among multiple motors on the same MT. In particular, for a MT velocity of $v = 50$nm/s, the longitudinal drag coefficient (per length) for aqueous solution $\Gamma_\parallel \approx 1.3 \times 10^{-3}$ Pa·s, and MT length $L = 2\mu m$, we would obtain a drag force on MTs of $f_{drag} = v L \Gamma_\parallel = 1 \times 10^{-4}$pN, which is orders of magnitude below relevant scales. We have assumed a processivity length of $l_{proc} = 800$ nm (100 steps with step length 8 nm) for the dimer motor constructs. In principle the monomer motor constructs should have a processivity of one step, but we have also considered a slightly larger processivity length (10 steps) motivated by the possibility that an effective processivity arises due to depletant-induced attractions between the motor constructs and MTs.

While the theory qualitatively captures the trends observed in the experiments, the discrepancies between the theoretical predictions and experimental results are enlightening. First, the theory predicts that velocities increase at the same value of overlap or monomeric and dimeric clusters, while experiments show that the dimeric clusters require larger overlaps. This discrepancy arises because in the theory we have assumed that the average force on motor constructs at a given overlap length is the same for monomeric and dimeric motor constructs, and thus the dimerized fraction is the same for both cases. We made this assumption because the dimerization fraction was not measured for the processive dimeric kinesin motor constructs, so it was not possible to determine an optimal average force $f^*$ for dimeric kinesin. However, we expect that the average force on motor constructs will be higher in the case of dimeric kinesin due to their higher processivity, as evidenced by increased maximum velocities for two-headed kinesin relative to single-headed kinesin. Second, the predicted decrease in velocity with overlap for low processivity is somewhat more gradual than the experimental observations for single headed kinesin. This could arise because we assume that the rate of motor construct reorientation is limited only by the kinesin stepping rate, whereas motor construct motions could be impeded by the dense milieu of the active nematic film.

**References for supplementary text**

*Illustrations were generated with Biorender.com

**Supplementary Figures**

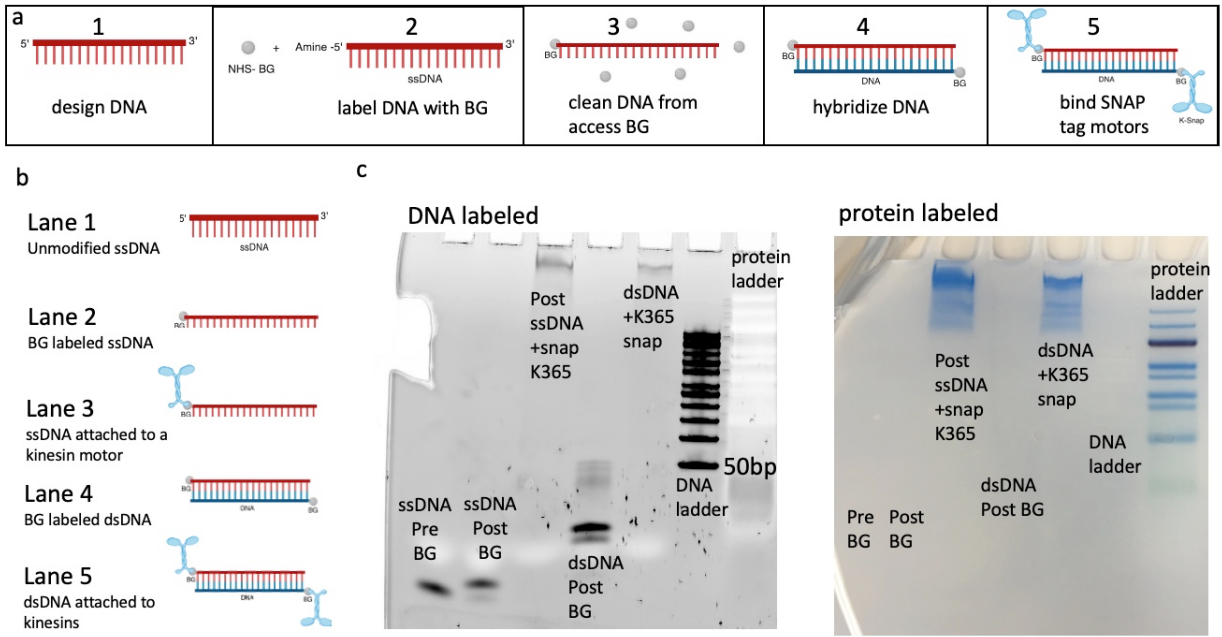

**Figure S1: (a)** Five steps assembly of DNA motor clusters. **(b)** DNA and protein constructs examined by DNA and protein gel. **(c)** Gel of SYBR gold DNA label gel (Left), and protein Coomassie blue dye (Right). Pre and post stand for DNA before and after BG modification. Gel is a 20% acrylamide TBE gel (ThermoFisher).

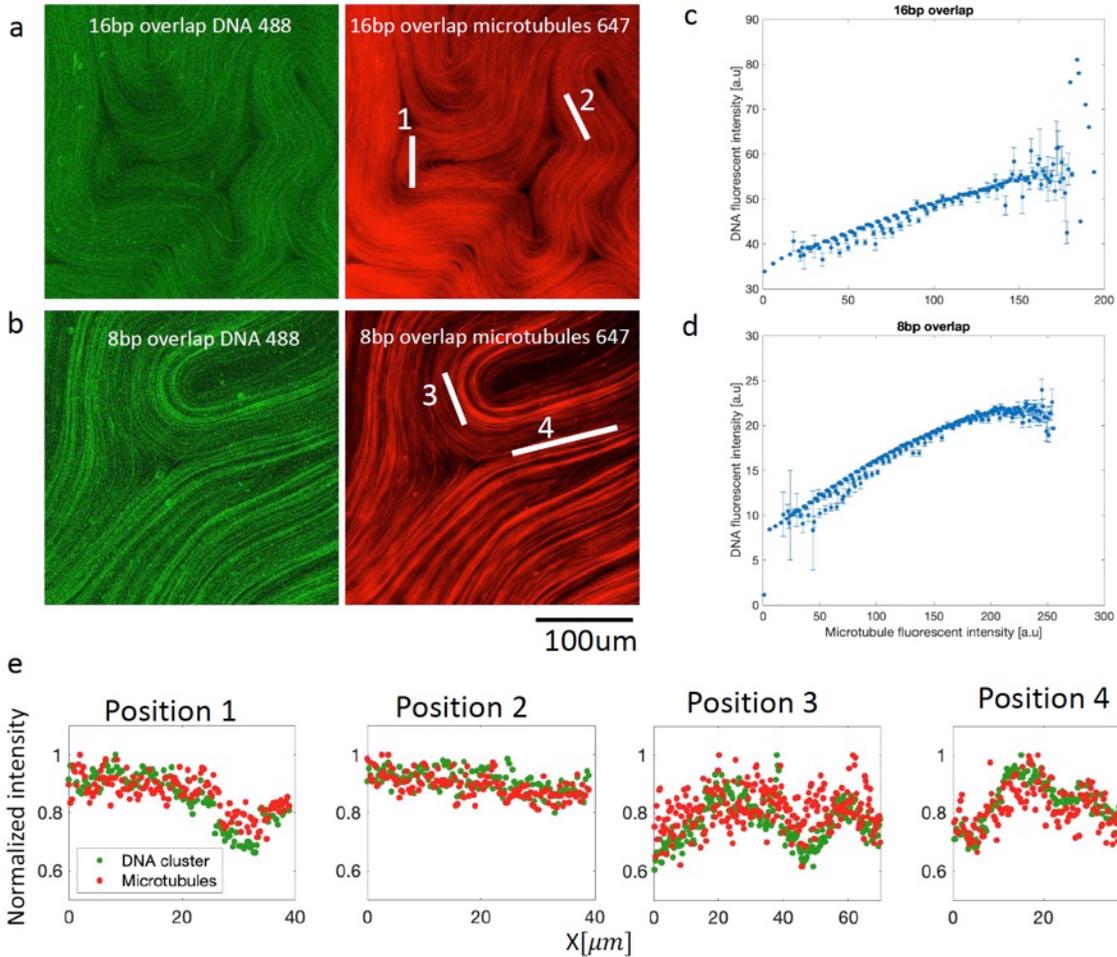

**Figure S2 – (a)** Fluorescent image of active nematics with single headed 16 bp overlap clusters labeled with SYBR green (green) and microtubules labeled with Alexa647 (red). **(b)** Equivalent image of active nematics that are powered by 8 bp overlap clusters. Scale bar, 100 $\mu m$. (c) DNA fluorescent signal scales linearly with the microtubule fluorescent signal for active nematics powered by16 bp overlap clusters. Each point is averaged over 1024 pixels and 50 time points. **(d)** Identical analysis extracted from active nematics powered by 8 bp overlap clusters. **(e)** Spatial profiles of normalized fluorescent intensity for both DNA and microtubules along the positions marked in (a) and (b).

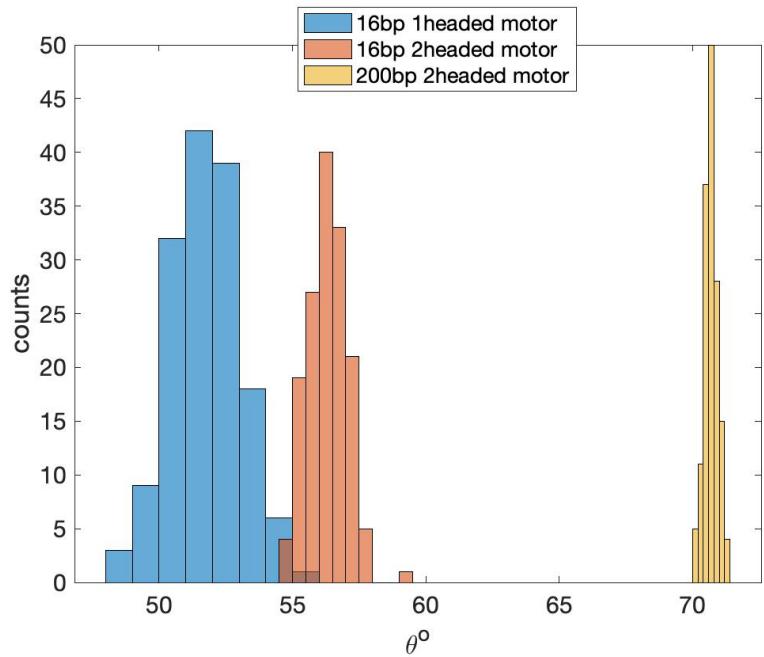

**Figure S3** – Distributions of the width of the orientational distribution function of motor clusters, $\theta$, calculated using bootstrapping. Data was subsampled 150 times to 50% of the data points. $\theta$ was repeatedly fitted using least squares regression. The generated distributions of theta are plotted for the different conditions in figure 2C. The mean values of the histograms are statistically different with p values <<0.001.

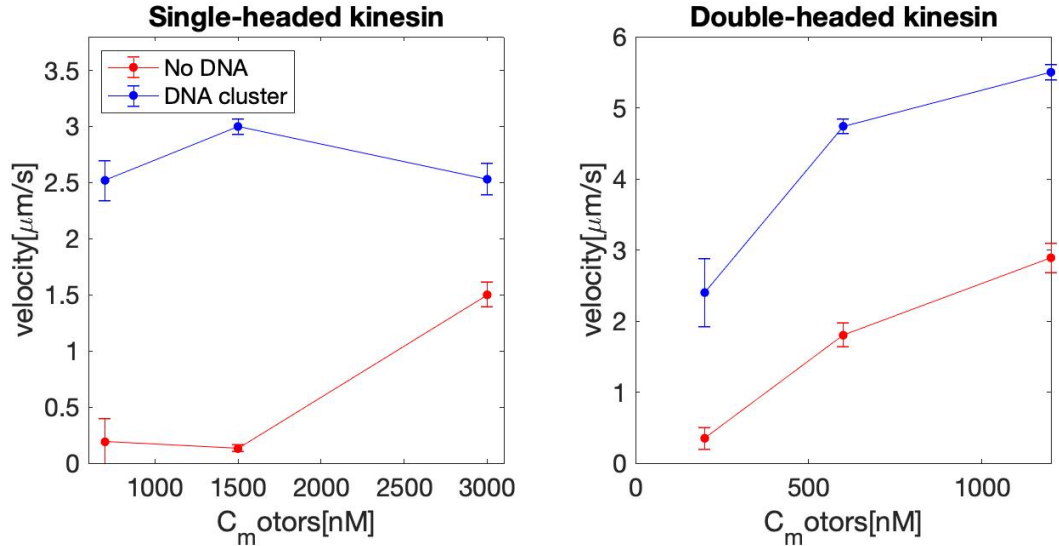

**Figure S4** – Active nematics velocity as a function of motor concentration for motors linked through DNA (blue) and motors without linking (red). **(a)** Measurements for a single headed non-processive kinesin motor, linker DNA with a 16bp overlap. **(b)** Measurements for a double headed processive kinesin motor, linker DNA 32bp overlap. Optimal working conditions are chosen such that the velocity range between the blue and red curves are maximal, while keeping minimal background activity. Error bars are standard errors averaged over 4 samples. DNA cluster concentration corresponds to 0.5 of the motor concentrations.

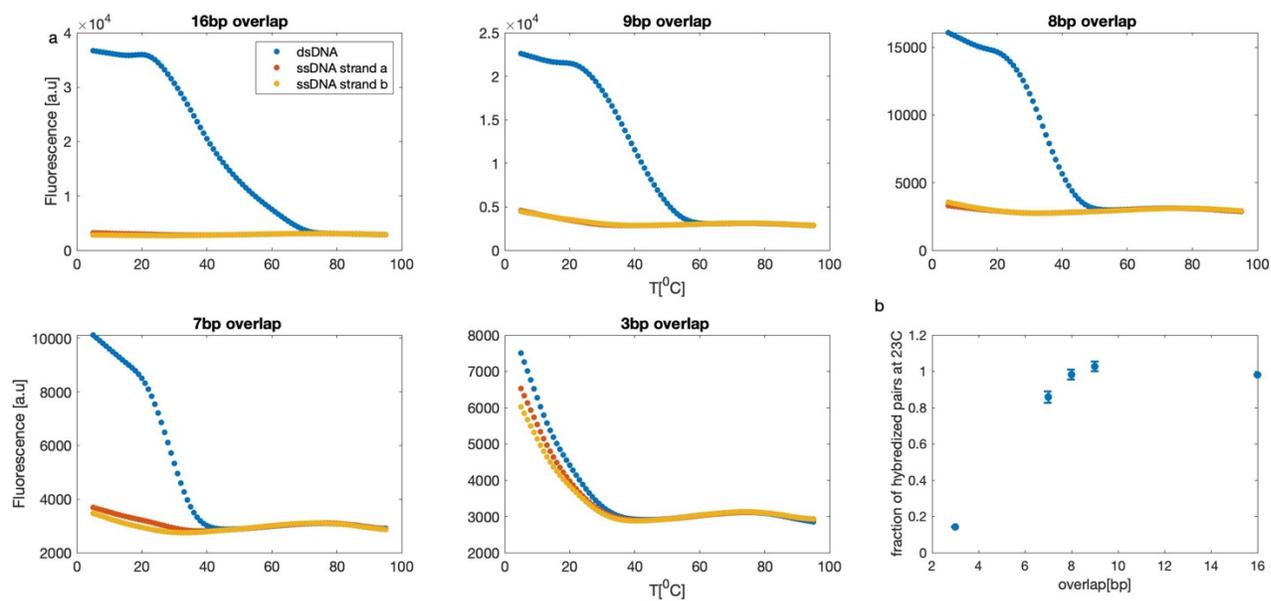

**Figure S5 - (a)** Melting curves for different DNA overlaps, measured using SYBR green with no activity in the sample. DNA concentration 250 nM. **(b)** Fraction of hybridized strands at 25 °C as a function of DNA overlap. Error bars are standard deviations of 3 independent measurements.

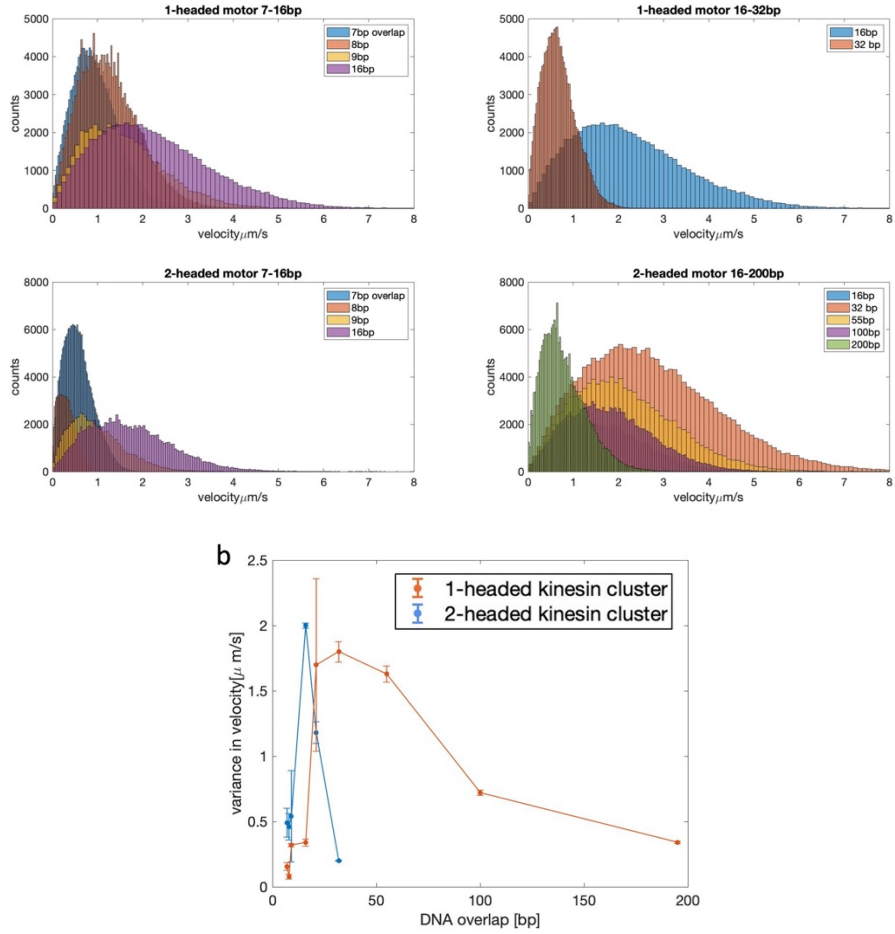

**Figure S6- (a)** Velocity distributions of active nematics powered by clusters of different overlap lengths. Each distribution is taken from a single data set. Velocities in the main text are time and spaces average of multiple separate repeats. **(b)** Variance of the velocity for different DNA overlaps. Variance follows the same trend as the velocities. Errors are taken as the standard error from 3 -8 separate repeats.

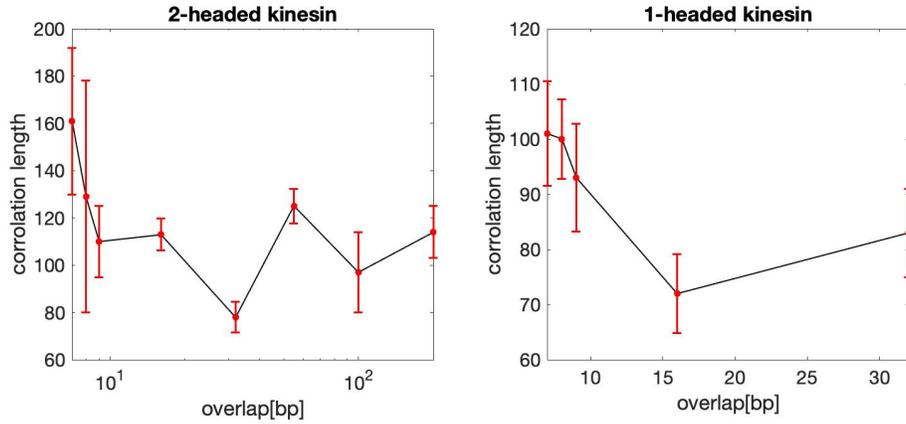

**Figure S7** - Length scale extracted from the spatial velocity-velocity correlation function as a function of DNA overlap, for both processive and non-processive clusters. The length scale was extracted from an exponential fit to velocity-velocity correlations analysis. Error bars are standard errors extracted from N=4-8 separate measurements.

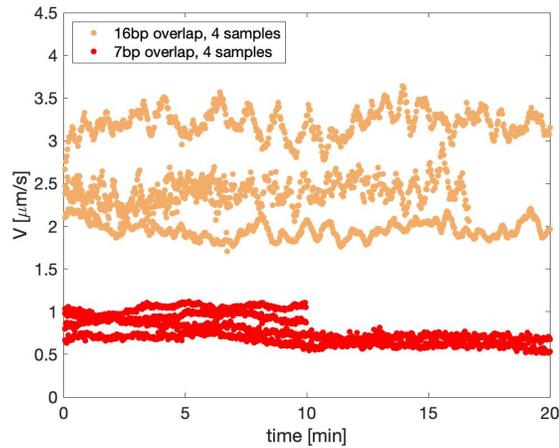

**Figure S8** – Speed of autonomous flows generated by active nematics as a function of time. Active nematics were powered by monomeric DNA motor cluster with 16 bp, and 7 bp overlaps.

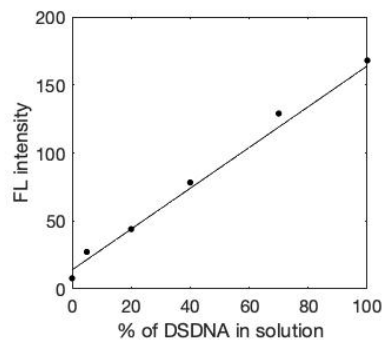

**Figure S9** - SYBR green signal as a function of double strand DNA in a buffer without activity. DNA concentration is kept constant at 350 nM. Double strand DNA is adjusted against single strand DNA. Data taken on a confocal microscope.

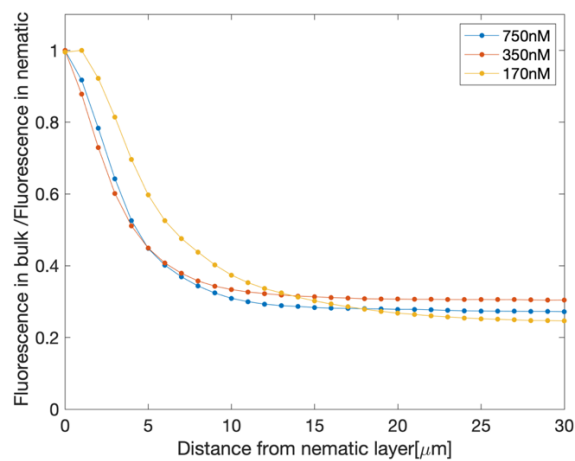

**Figure S10** - Fluorescent intensity z-profiles of non-processive cluster with a 16 bp overlap as a function of the distance from the 2D nematic layer. Clusters were labeled with SYBR green to only account only for the ds-DNA.

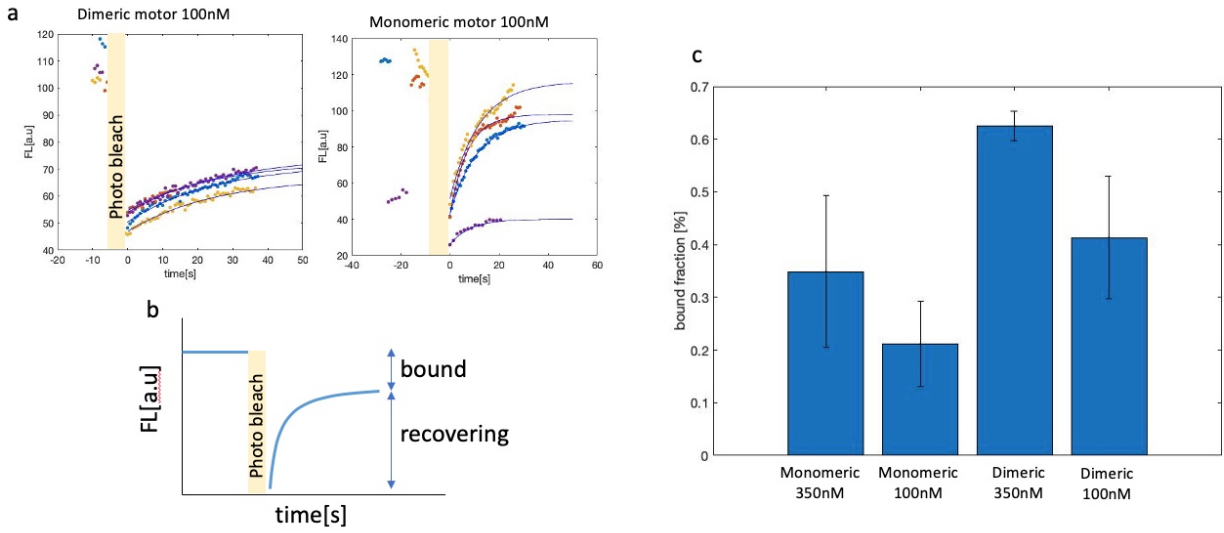

**Figure S11 - (a)** Fluorescence recovery of processive and non-processive cluster within active nematics after photobleaching. **(b)** The fluorescence does not recover to the pre-bleached intensity levels. Schematic showing the estimate of the fraction of permanently bound and recovering motors. **(c)** Fraction of motors remaining bound after recovery from photobleaching. This could be due to rigor motors, damage from photo bleaching and/or difference in processivity.

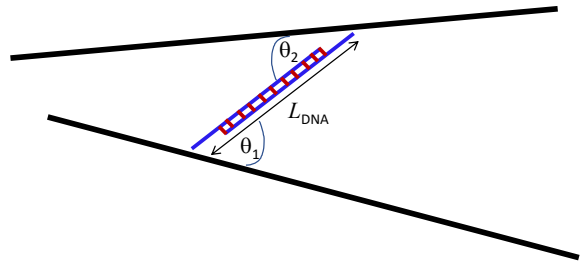

**Figure S12** - Schematic of the model for motor construct reorientation. The motor construct (twin blue lines, with red lines indicating base pairs) binds between two microtubules (thick black

lines). We take the initial angle of the motor construct relative to the microtubules as the mean of the angles at either end: $\theta_0 = (\theta_1 + \theta_2) / 2$.